\documentclass[runningheads]{llncs}
\usepackage{ amssymb }
\usepackage{graphicx}
\usepackage{soul,color}
\usepackage{chapterbib}
\usepackage{ amssymb }
\usepackage{graphicx}
\usepackage{soul,color}
\usepackage{multirow}
\usepackage{hyperref}
\def\ie{\emph{i.e}}

\begin{document}
\title{KiU-Net: Towards Accurate Segmentation of Biomedical Images using Over-complete Representations}
%
%
\author{Jeya Maria Jose Valanarasu\inst{1}\and
	Vishwanath A. Sindagi\inst{1} \and
	Ilker Hacihaliloglu\inst{2} \and
	Vishal M. Patel\inst{1}}


\authorrunning{JMJ Valanarasu et al.}
%
\institute{Johns Hopkins University, Baltimore, MD, USA \and
	Rutgers, The State University of New Jersey, NJ, USA 
}
\maketitle              
\begin{abstract}
Due to its excellent performance, U-Net is the most widely used backbone architecture for biomedical image segmentation in the recent years.  However,  in our studies, we observe that there is a considerable performance drop in the case of detecting smaller anatomical structures with blurred noisy boundaries. We analyze this issue in detail, and address it by proposing an over-complete architecture (Ki-Net) which involves projecting the data onto  higher dimensions (in the spatial sense). This network, when augmented with  U-Net, results in significant improvements in the case of segmenting small anatomical landmarks and blurred noisy boundaries while obtaining better overall performance. Furthermore, the proposed network  has additional benefits like faster convergence and fewer number of parameters. We evaluate the   proposed method on the task of brain anatomy segmentation from 2D Ultrasound (US) of preterm neonates, and achieve an improvement of around $4\%$ in terms of the DICE accuracy and Jaccard index as compared to the standard-U-Net, while outperforming the recent best methods  by $2\%$. Code: \href{https://github.com/jeya-maria-jose/KiU-Net-pytorch}{https://github.com/jeya-maria-jose/KiU-Net-pytorch}


\end{abstract} 

\section{Introduction}

Preterm birth is among the leading public health problems in the USA and Europe \cite{ment2009imaging}. The reported annual cost of care for preterm neonates exceeds \$18 billion dollars every year in the USA alone \cite{ment2009imaging}. Although, advancements made in neonatal care have increased the survival rates, majority of these infants are at risk for adverse neuro-developmental outcomes. Among the different types of preterm brain injury, intraventricular hemorrhage (IVH) remains the most common cause of acquired hydrocephalus resulting in the enlargement of ventricles. On the other hand, absence of septum pellucidum is used as a biomarker for the diagnosis of other brain disorders such as septo-optic dysplasia. Cranial ultrasound (US) remains the main imaging modality used to diagnose brain disorders in preterm neonates due to its real-time, safe, and cost effective imaging capabilities. Current clinical evaluation involves qualitative investigation of the collected US scans or quantitative manual measurement of landmarks such as ventricular index (VI), anterior horn width (AHW), frontal and temporal horn ratio (FTHR) \cite{el2010neuroimaging}. Qualitative evaluation is subjective and manual measurement involves intra and inter-user variability errors. The diagnostic accuracy is further affected by the unclear boundary of the ventricles, due to build up of bleeding pressure, or sub-optimal orientation of the transducer during imaging. Additionally, shading artifacts  causes incomplete boundaries in the acquired US data. Depending on the bleeding extend, the shape of the ventricle varies for different subjects. Finally, manual measurement is also problematic for normal preterm neonates without any brain injury due to very small ventricle size and blurred boundaries. Similar problems are also faced for identifying septum pellucidum due to its small size and unclear boundary. In order to overcome these challenges, precise and automatic segmentation of ventricles and septum pellucidum is critical for accurate diagnosis and prognosis.

Several groups have proposed semi-automatic and fully automatic methods for segmentation of ventricles from 2D/3D US scans. Methods based on traditional medical image analysis are time consuming or not robust enough to the previously mentioned challenging scan conditions \cite{boucher2018dilatation,tabrizi2018automatic,qiu2017automatic}. The reported DICE similarity coefficient values were $70.8\%$ \cite{boucher2018dilatation}, $80\%$ \cite{tabrizi2018automatic}, and $76.5\%$ \cite{qiu2017automatic}. The reported computation times were 54 minutes for \cite{qiu2017automatic}. The other methods did not report any computation time. Most recently, methods based on deep learning were also investigated by various groups to improve the robustness and computation time of segmentation \cite{martin2018automatic,wang2018automatic,valanarasu2020learning}. Since the introduction of U-Net \cite{ronneberger2015u} in 2015, it has been the leading deep learning-based network of any method that deals with biomedical image segmentation \cite{cciccek20163d,milletari2016v,zhao2019fully,li2018h,islam2019brain,islam2018ischemic,zhou2019unet++}. In \cite{martin2018automatic}, a U-Net architecture was used for segmentation of ventricles. 

\begin{figure}[t!]
	\centering
	\includegraphics[width=.9\linewidth]{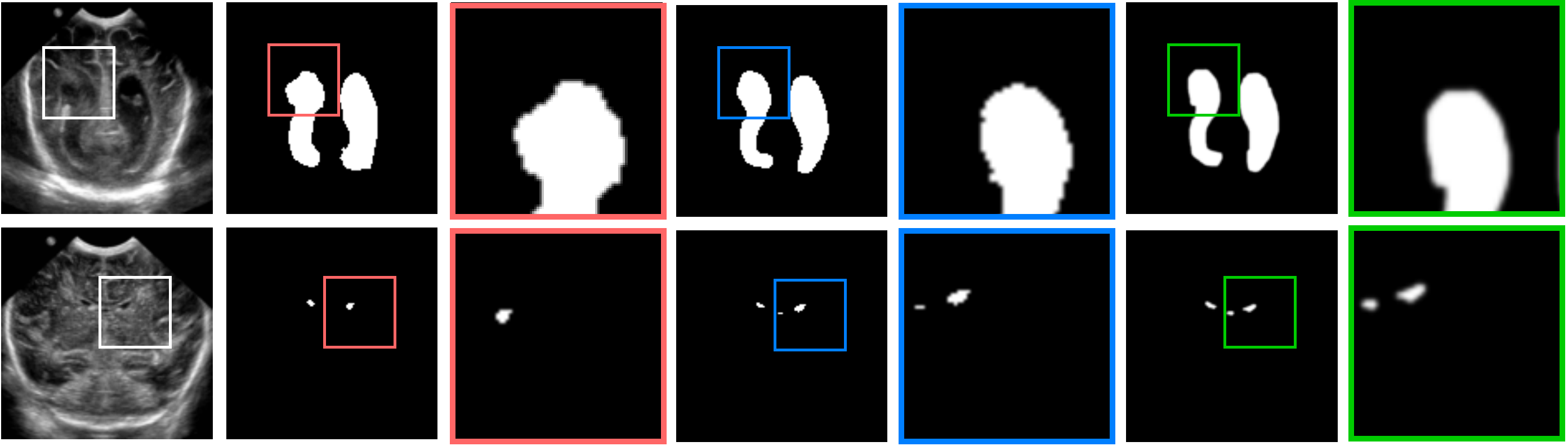}\\
		(a) \hskip30pt (b) \hskip30pt (c) \hskip30pt (d) \hskip30pt (e) \hskip30pt (f) \hskip30pt (g)\\
\vskip -10pt	\caption{(a) Input B-Mode Ultrasound Image. Predictions from (b) U-Net, (d) KiU-Net (ours), (f) Ground Truth. (c),(e) and (g) are the zoomed in patches from (b),(d) and (f) respectively. The boxes in the original images correspond to the zoomed in portion for the zoomed images. It can be seen that our proposed network captures edges and small masks better than U-Net. }
	
	\label{initvis}
\end{figure}

Based on the observations  that the existing approaches do not achieve optimal performance (especially in the case of segmenting out small anatomical structure), we analyze this issue in detail. Specifically, we conducted experiments with the standard U-Net architecture which is a leading backbone in several segmentation algorithms. In spite of the skip connections that enable the propagation of information from shallower layers to deeper layers, the network is unable to capture finer details (see Fig. \ref{initvis}) for the following reasons. The standard encoder-decoder architecture of U-Net belongs to the family of  under-complete convolutional autoencoders, where the dimensionality of data is reduced near the bottleneck. The initial few blocks of the encoder learn low-level features of the data while the later blocks learn the high-level features. Eventually, the encoder learns to map the data to lower dimensionality (in the spatial sense). The increasing receptive field size  over the depth of the network, constrains the network to focus more on the higher-level features. However, it is important to note that tiny structures require smaller receptive fields.  In the case of  standard U-Net, even with skip connections,  the smallest receptive field is limited by that of the first layer. Hence, under-complete architectures are essentially limited in their abilities to capture finer details.

Considering the aforementioned drawback of under-complete representations, we resort to over-complete architectures where the data is projected onto a higher dimension in the intermediate layers. In the literature, over-complete representations have been shown to be more robust and  stable, especially in the presence of noise \cite{lewicki2000learning}. However, such architectures have been  relatively unexplored for segmentation tasks in both the computer vision and medical imaging communities \cite{haque2020deep}. In this paper, we explore the use of such an over-complete network for segmentation to address the issue of lack of smaller receptive field in the standard U-Net. 
We refer to the over-complete network as Kite-Net (Ki-Net) as it's shape is similar to that of a kite. In the following sections, we show how the information learned by Ki-Net actually helps in capturing finer shape structures and edges better than the generic under-complete networks. Furthermore, we propose  to effectively combine the benefits of the proposed Ki-Net with that of the standard U-Net using a novel cross-scale fusion strategy. We show that this novel network (KiU-Net) achieves state-of-the-art performance on the brain anatomy segmentation task from US images when compared with the latest methods.

In summary, this paper (1) explores over-complete deep networks (Ki-Net) for the task of  segmentation, (2) proposes a novel architecture (KiU-Net) combining the features of both under-complete and over-complete deep networks which captures finer details better than the standard encoder-decoder architecture of U-Net thus aiding in  precise segmentation, and (3) achieves faster convergence and better performance metrics than recent methods for segmentation.


\section{Proposed Method}

\begin{figure}[t!]
	\centering
	\includegraphics[width=.6\textwidth]{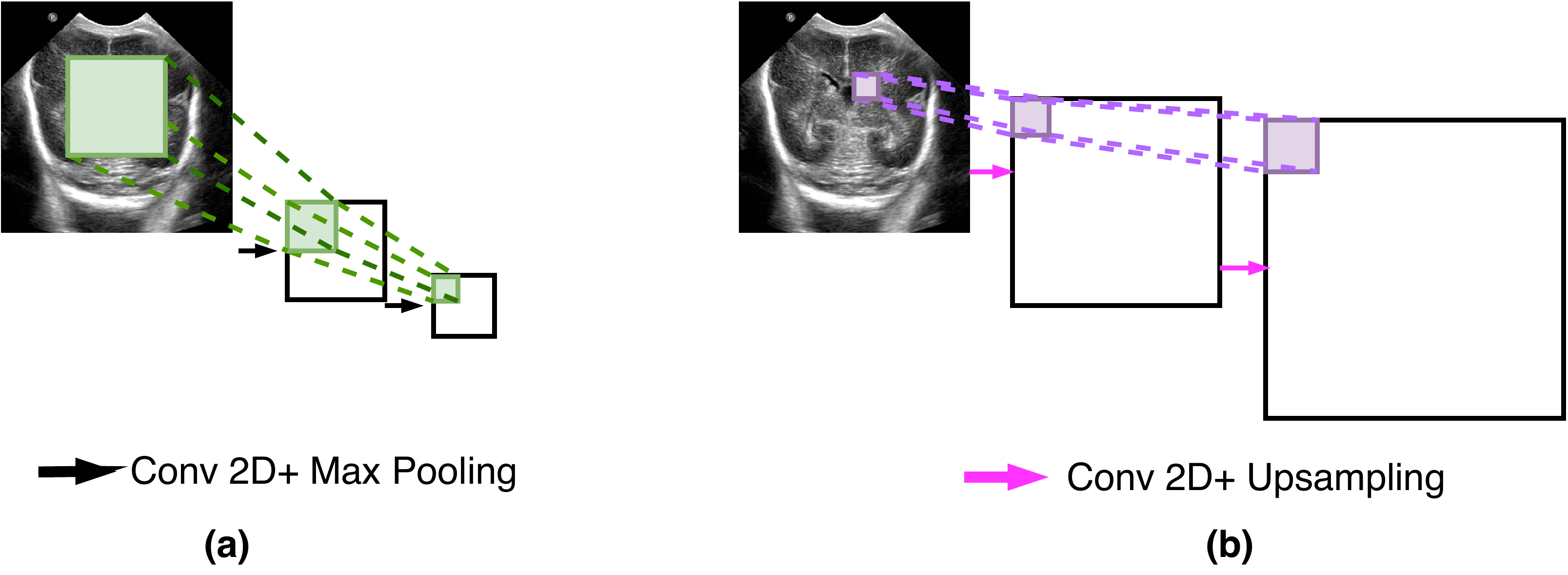}
	\vskip -15pt
	\caption{Effect of architecture type on receptive field. (a) U-Net: Each location in the intermediate layers focuses on a much larger region in the input. (b) Ki-Net: Each location in the intermediate layers focuses on a much smaller region in the input. }
	\label{fig:explain}
\end{figure}

\vskip -10pt\noindent {\bf{Over-complete representations: }}
As illustrated in Fig  \ref{fig:explain},  the receptive field of the filters in a generic ``encoder-decoder" architecture increases as we go deeper in the network. This increase in receptive field size can be attributed to two reasons: (i) every conv layer filter gathers information from a surrounding window, and (ii) the use of max-pooling layer after every conv layer. The max-pooling layers  essentially double  the receptive field size after every conv layer. The increasing receptive field reasons is critical for CNNs to learn high-level features like objects, shapes or blobs. However, a side effect of this is that  it reduces the focus of the filters. That is, except the first layer, filters in the other layers have reduced abilities  to learn features that correspond to fine details like edges and their texture. This causes any network with the standard under-complete architecture to not produce sharp predictions around the edges in tasks like segmentation. 

To overcome this issue, we propose Ki-Net which is over-complete in the spatial sense. That is, the spatial dimensions of the intermediate layers is more than that of the input data. We achieve this by employing an upsampling layer  after every conv layer  in the encoder. Furthermore, we employ a max pooling layer after every conv layer in the decoder in order to reduce the dimensionality back to that of the input. This forces the over-complete conv architecture to behave differently than the standard under-complete conv architecture. The filters in this type of architecture learn finer low-level features due to the decreasing size of receptive field even as we go deeper in the encoder network. 

\begin{figure}[t!]
	\centering
	\includegraphics[width=.8\linewidth]{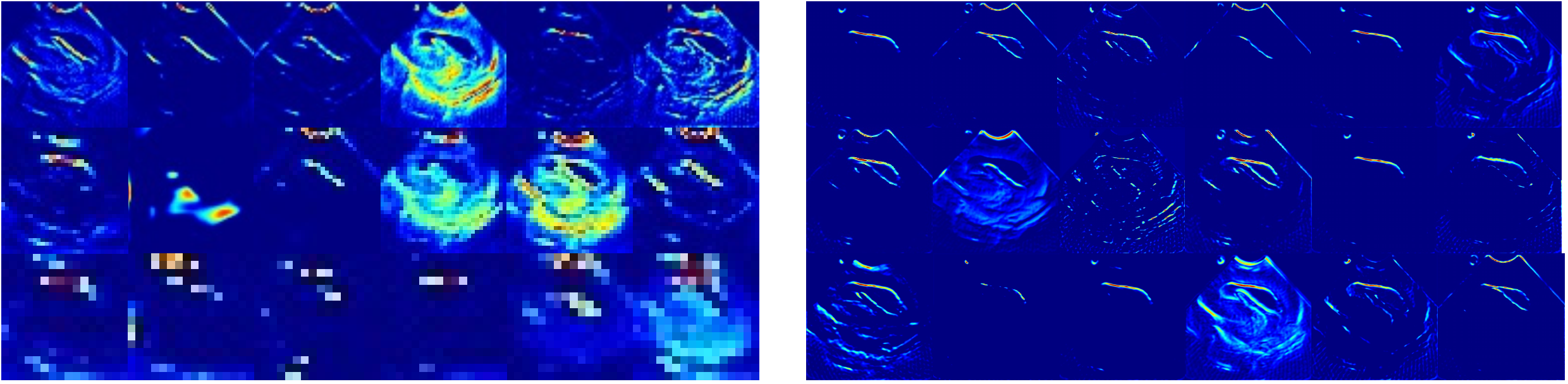}\\
	(a)\hskip140pt(b)
	\vskip-12pt
	\caption{Visualization of filter responses for (a) U-Net, and (b) Ki-Net. Top row: Feature maps from the first layer of encoder. Middle row: Feature maps from the second layer of encoder. Bottom row: Feature maps from the third layer of encoder. By restricting the receptive field, Ki-Net is able to focus on edges and smaller regions.}
	
	\label{aefil}
\end{figure}

Fig \ref{fig:explain}(a) illustrates  how the receptive field is large for U-Net. Fig \ref{fig:explain}(b) illustrates how the use of over-complete architecture like Ki-Net restricts the receptive field size to a smaller region. Hence, by constricting the receptive field size, we force the filters in the deeper layers to learn very fine edges as it tries to focus heavily on smaller regions. To illustrate this, we show how the filters of encoder fire in a Ki-Net when compared to U-Net in Fig \ref{aefil}. It can be observed that the filters in U-Net become smaller as we go deeper and fire across high-level shapes where as the filters become bigger as we go deeper in Ki-Net and the features captured are fine edges across all layers with an increased resolution.


\begin{figure}[t!]
	\centering
	\includegraphics[width=.8\linewidth]{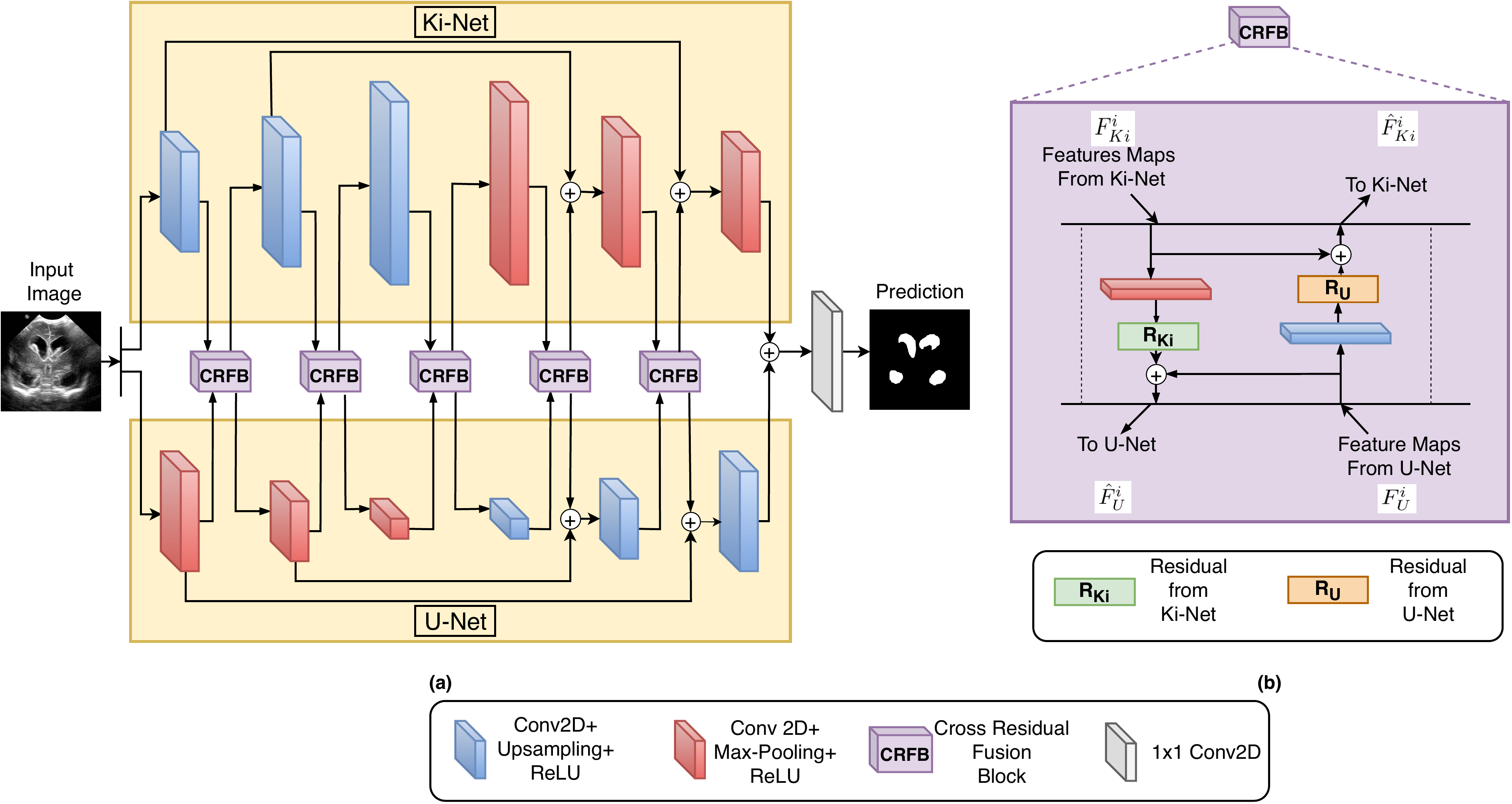}
	\vskip -10pt
	\caption{(a) An overview of the  proposed KiU-Net architecture. (b) Cross Residual Fusion block architecture. }
	\label{Ki-Netplus}
\end{figure}

\noindent {\bf{KiU-Net: }}
As we have established that our proposed Ki-Net has better abilities to captures edges compared to U-Net, we combine it with the standard U-Net in order to improve the overall segmentation accuracy as Ki-Net if used separately will only capture the edges. The combined network, KiU-Net, exploits the low-level fine edges capturing feature maps of Ki-Net as well as the high-level shape capturing feature maps of U-Net. We propose using a parallel network architecture where one branch is a Ki-Net and the other a U-Net as seen in Figure \ref{Ki-Netplus}(a). The input image is forwarded through both the branches simultaneously. In both the branches, we have 3 layers of conv blocks in the encoder as well as the decoder. Each conv block in the encoder of Ki-Net branch consists of a 2D conv layer followed by a bilinear interpolation with a scale factor of $2$ and ReLU non-linearity. Similarly, each conv block  in the decoder of Ki-Net branch consists of a 2D conv layer followed by a max-pooling layer with a pooling coefficient of two. In addition, we use skip connections between the blocks of encoder and decoder similar to U-Net to enhance the localization.  In the U-Net branch, we adopt the ``encoder-decoder" architecture of a U-Net. 

In order  to augment the two networks, one can perform    simple concatenation of features at the final layer. However, this may not be necessarily optimal. Instead, we combine the feature maps at each block and this  results in better convergence as the flow of gradients during back propagation is across both the branches at each block level \cite{sindagi2019multi}. Furthermore, in order to combine the features at each block level more effectively, we propose a cross residual fusion block (CRFB). This block  extracts complementary features from both network branches and forwards to both of them respectively. Specifically, the CRFB consists of residual connections, followed by a set of conv layers (see Fig. \ref{Ki-Netplus} (b)). In order to combine the feature maps from the two networks $F_{U}^i$ (U-Net) and  $F_{Ki}^i$  (Ki-Net) after the $i^{th}$ block, cross-residual features $R_U^i$ and $R_{Ki}^i$ are first estimated through a set of conv layers. These cross-residual feature are then added to the original features $F_{U}^i$ (U-Net) and  $F_{Ki}^i$   to obtain the complementary features $\hat{F}_{U}^i$ and  $\hat{F}_{Ki}^i$, \ie,  $\hat{F}_{U}^i = F_{U}^i + R_{Ki}^i$ and $\hat{F}_{Ki}^i = F_{Ki}^i + R_{U}^i$. This strategy is more effective compared to simple feature fusion schemes like addition or concatenation. Finally, the features from  decoder in both the branches are added and forwarded through \(1\times1\) conv layer to produce the final segmentation mask. The complete details of the network such as the kernel size, number of filters, etc. are included in supplementary material.


We train the network using  pixel-wise binary cross entropy loss  between the prediction and ground-truth. The loss function between the  prediction $p$ and the ground truth $\hat{p}$  is defined as follows: 
	\setlength{\belowdisplayskip}{0pt} \setlength{\belowdisplayshortskip}{0pt}
\setlength{\abovedisplayskip}{0pt} \setlength{\abovedisplayshortskip}{0pt}
\[\mathcal{L}_{CE(p,\hat{p})} = - \frac{1}{wh} \sum_{x=0}^{w-1}\sum_{y=0}^{h-1}(p(x,y) \log(\hat{p}(x,y)) ) + (1-p(x,y))log(1-\hat{p}(x,y)),\]
where $w$ and $h$ are the dimensions of image, $p(x,y)$ and  $\hat{p}(x,y)$ denote the output at a specific location $(x,y)$ of the prediction and ground truth, respectively. 

\section{Experiments and results}


\vskip-10pt\noindent {\bf{Dataset acquisition and details: }}
After obtaining institutional review board (IRB) approval, US scans were collected from 20 different premature neonates (age $<1$ year). The dataset contains subjects with IVH as well as healthy ones. The US scans were collected using a Philips US machine (Philips iE33) with a C8-5 broadband curved array transducer using coronal and sagittal scan planes. Imaging depth and resolution varied between 6-8 cm and 0.1-0.15 mm, respectively. Ventricles and septum pellecudi were manually segmented by an expert ultrasonographer. A total of 1629 images with annotations were obtained in total. The scans were randomly divided into 1300 images for training and 329 images for testing. This process was repeated 3 times. During random split the training and testing data did not include scans from the same patient. Before processing the resolution of each image was changed to $128\times 128$.

\noindent {\bf{Implementation details: }}  KiU-Net is trained using  cross-entropy loss $\mathcal{L}_{CE(p,\hat{p})}$ with the Adam optimizer \cite{kingma2014adam} and a batch-size of 1. The learning rate was set equal to 0.001. The network was built in PyTorch framework and trained using Nvidia-RTX 2080Ti GPUs. The network was trained for a total of 100 epochs. 
 
\noindent {\bf{Comparison with recent methods: }}
Since the main focus of this work is to augment the U-Net architecture with additional capabilities, we compare our method   with U-Net and  other recent methods. Table \ref{Comp} shows that  the proposed method performs better than other recent methods  like Seg-Net \cite{badrinarayanan2017segnet}, pix2pix \cite{isola2017image}, and Wang et al. \cite{wang2018automatic}. Seg-Net \cite{badrinarayanan2017segnet} has been most recently investigated for segmentation of kidneys from US data \cite{yin2020automatic}, pix2pix \cite{isola2017image} has been used for multi-task organ segmentation from chest x-ray radiography\cite{eslami2019image}, and Wang \textit{et al.} \cite{wang2018automatic} has been previously used for segmentation of ventricles from brain US data. We run the experiments 3 times for different random folds of training and testing data and report the mean metrics with the variance. 

It can be observed that the proposed method  achieves an  improvement of $4\%$ in DICE accuracy with respect to U-Net and a $2\%$ improvement with respect to state-of-the-art \cite{wang2018automatic} (see Table \ref{comp}). Fig. \ref{Fig:exp} illustrates the prediction of segmentation masks using different methods along with the input and ground truth. From the first row in Fig. \ref{Fig:exp}, we can observe that KiU-Net (our method) is able to predict even very small masks precisely, whereas all the other methods fail. Similarly, from the second row we can observe that our network detects the edges better than other methods. This demonstrates that the intuition of constricting the receptive field size by following the over-complete representation served its purpose as the smaller masks are not missed in our method. Additionally, it may be noted that the proposed method performs well irrespective of the size of the anatomy structures.   Furthermore, the proposed network has the following additional benefits. First, it uses much fewer number of parameters in comparison to the other methods. Note that U-Net used in KiU-Net has less number of blocks and filters compared to the original U-Net as in \cite{ronneberger2015u}, thus resulting in less number of parameters. Second, it converges much faster compared to the standard U-net (see Fig. 6). Its inference time is 8 ms for one test image.
 
\vskip -15pt
\begin{table}[htp!]
	\begin{center}
		\centering
		\caption{Comparison of results. Proposed method outperforms existing approaches.}
		\vskip -10pt
		\resizebox{0.7\textwidth}{!}{
			\label{Comp}
			\begin{tabular}{l|c|c|c}
				\hline
				Method       & DICE Acc (\%)  & Jaccard Idx (\%)  & Parameters\\ \hline 
				Seg-Net \cite{badrinarayanan2017segnet}  & 82.79 $\pm$ 0.320   & 75.02 $\pm$ 0.570   & 12.5M      \\ 
				
				U-Net \cite{ronneberger2015u}        & 85.37 $\pm$ 0.002  &  79.31 $\pm$ 0.065 & 3.1M  \\
				
				pix2pix \cite{isola2017image}        & 85.46 $\pm$ 0.022 & 77.45 $\pm$ 0.56 &54.4M    \\ 
				
				Wang et.al\cite{wang2018automatic}   & 87.47 $\pm$ 0.080    & 80.51 $\pm$ 0.190 & 6.1M      \\ 
				
				KiU-Net (ours) & \textbf{89.43 $\pm$ 0.013 }  & \textbf{83.26 $\pm$ 0.047} & \textbf{0.29M}  \\ \hline
			\end{tabular}
			
			\label{comp}	
		}
	\end{center}
\end{table}

\begin{figure}[t!]
	\begin{center}
		\centering
				\includegraphics[width=0.125\textwidth,height = 0.125\textwidth]{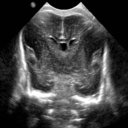}
		\includegraphics[width=0.125\textwidth,height = 0.125\textwidth]{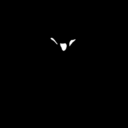}
		\includegraphics[width=0.125\textwidth,height = 0.125\textwidth]{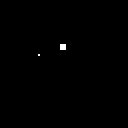}
		\includegraphics[width=0.125\textwidth,height = 0.125\textwidth]{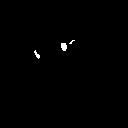}
		\includegraphics[width=0.125\textwidth,height = 0.125\textwidth]{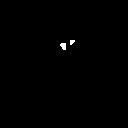}
		\includegraphics[width=0.125\textwidth,height = 0.125\textwidth]{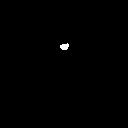}
		\includegraphics[width=0.125\textwidth,height = 0.125\textwidth]{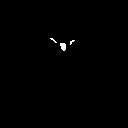}

		%
		
		\vskip4pt
		\includegraphics[width=0.125\textwidth,height = 0.125\textwidth]{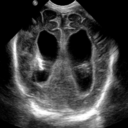}
		\includegraphics[width=0.125\textwidth,height = 0.125\textwidth]{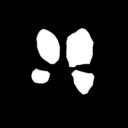}
		\includegraphics[width=0.125\textwidth,height = 0.125\textwidth]{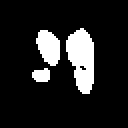}
		\includegraphics[width=0.125\textwidth,height = 0.125\textwidth]{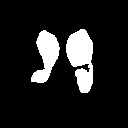}
		\includegraphics[width=0.125\textwidth,height = 0.125\textwidth]{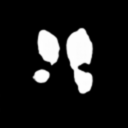}
		\includegraphics[width=0.125\textwidth,height = 0.125\textwidth]{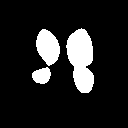}
		\includegraphics[width=0.125\textwidth,height = 0.125\textwidth]{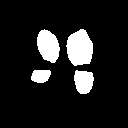}
		
		
		
		(a)\hskip35pt(b)\hskip35pt(c)\hskip35pt(d)\hskip35pt(e)\hskip35pt(f)\hskip35pt(g)\\
		\vskip -10pt
		\caption{Qualitative results on sample test images. (a) B-mode input US image. (b) Ground truth.  (c) Seg-Net \cite{badrinarayanan2017segnet}. (d) U-Net \cite{ronneberger2015u} (e) pix2pix \cite{isola2017image}. (f) Wang et al. \cite{wang2018automatic}. (g) KiU-Net (ours).}
		\label{Fig:exp}
	\end{center}
\end{figure}

\begin{figure}[htp!]
	\centering
	
	\begin{minipage}{0.5\textwidth}
		\includegraphics[width=.75\linewidth]{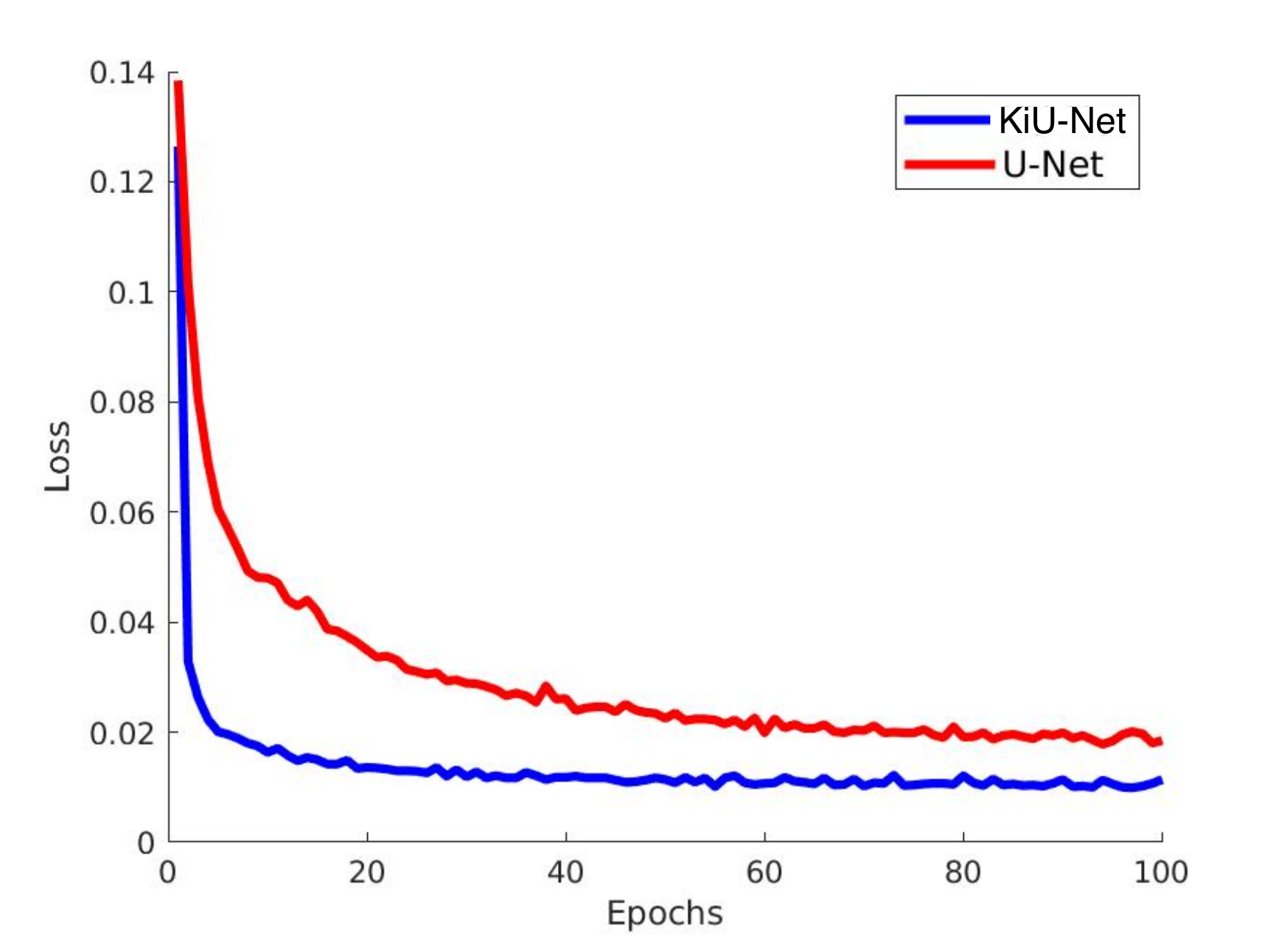}
		\vskip -12pt\caption{Comparison of convergence of the loss between KiU-Net and U-Net.}
	\end{minipage}\hskip25pt
	\begin{minipage}{.15\textwidth}
		\resizebox{2\linewidth}{!}{
			
			\begin{tabular}{l|c|c}
				\hline
				Method       & DICE   & Jaccard   \\ \hline 
				UC         &  82.79   & 75.02     \\
				OC         & 56.04   &  43.97    \\ 
				OC+UC     &  84.80     &   76.48          \\
				
				UC with SK         &  85.37  &   79.31    \\
				
				OC with SK         & 60.38 &  47.86    \\
				
				OC+UC  with SK   & 86.24      &  78.11        \\ 
				
				KiU-Net (ours) & \textbf{89.43 }  & \textbf{83.26}   \\ \hline
				
			\end{tabular}
			\label{abl}
		}
	\vskip -10pt	\caption{Ablation study.}
	\end{minipage}\hskip25pt
	\label{loss}
	
\end{figure}

\begin{figure}[htp!]
	\begin{center}
		\centering
		\includegraphics[width=0.125\textwidth,height = 0.125\textwidth]{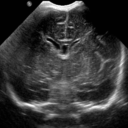}
		\includegraphics[width=0.125\textwidth,height = 0.125\textwidth]{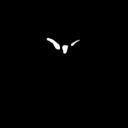}
		\includegraphics[width=0.125\textwidth,height = 0.125\textwidth]{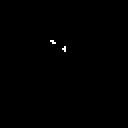}
		\includegraphics[width=0.125\textwidth,height = 0.125\textwidth]{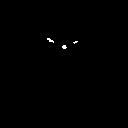}
		\includegraphics[width=0.125\textwidth,height = 0.125\textwidth]{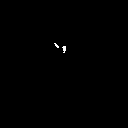}
		\includegraphics[width=0.125\textwidth,height = 0.125\textwidth]{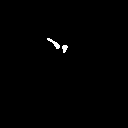}
		\includegraphics[width=0.125\textwidth,height = 0.125\textwidth]{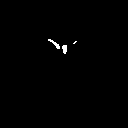}

		\vskip4pt
		\includegraphics[width=0.125\textwidth,height = 0.125\textwidth]{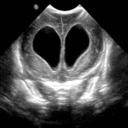}
		\includegraphics[width=0.125\textwidth,height = 0.125\textwidth]{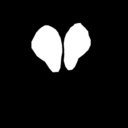}
		\includegraphics[width=0.125\textwidth,height = 0.125\textwidth]{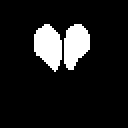}
		\includegraphics[width=0.125\textwidth,height = 0.125\textwidth]{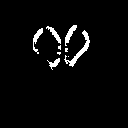}
		\includegraphics[width=0.125\textwidth,height = 0.125\textwidth]{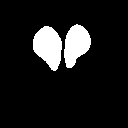}
		\includegraphics[width=0.125\textwidth,height = 0.125\textwidth]{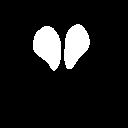}
		\includegraphics[width=0.125\textwidth,height = 0.125\textwidth]{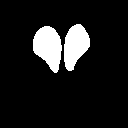}

		
		(a)\hskip35pt(b)\hskip35pt(c)\hskip35pt(d)\hskip35pt(e)\hskip35pt(f)\hskip35pt(g)\\
		\vskip -10pt
		\caption{Qualitative results of ablation study on test images. (a) B-Mode input US image. (b) Ground Truth annotation. Prediction of segmentation masks by (c) UC - Under-complete architecture (d) OC - Over-complete architecture (e) UC + SK (under-complete architecture with skip connections) (f) UC + OC with SK (combined architecture with skip connections) (g) KiU-Net (ours)}
		\label{Fig:abl}
	\end{center}
\end{figure}

\vskip-25pt\noindent {\bf{Ablation study: }}
We study the performance of each block's contribution to our KiU-Net by conducting a detailed ablation study. The results are shown in Fig 7. We start with the standard under-complete architecture (UC) and the over-complete architecture (OC). It can be noted here that the performance of OC is lesser than UC because even though OC captures the edges properly it does not capture most high level features like UC. Then, we show that  fusing both the networks (OC+UC) just by combining the feature maps at the final layer helps in improving the performance. This is followed by an experiment where we use skip connections (SK). It may be noted that UC with SK is basically the U-Net. Finally, we incorporate the cross residual fusion block (CRFB) at each block level  in our KiU-Net,  resulting in further improvements which demonstrates the effectiveness of our novel cross fusion strategy.  Fig \ref{Fig:abl} illustrates the qualitative improvements after adding each major block.

\vskip -25pt More results  on different datasets can be found in supplementary material. 

\section{Conclusion}
We  proposed a novel network called KiU-Net which is constructed by  augmenting the standard under-complete architecture based U-Net with an over-complete structure (Ki-Net). The purpose of Ki-Net is to specifically capture fine edges and small anatomical structures which are typically missed out in the other methods.  Further, we incorporate a new fusion strategy that is based on cross-scale residual blocks which results in a more effective use of information from the two networks. The proposed network has additional benefits like it uses much fewer number of parameters and results in faster convergence. The proposed method achieves better performance  as compared to recent methods on a relatively complex dataset which has both small and big segmentation masks. 

\section*{
	\centering Appendix}

\section*{Experiments on other modalities}
In the paper, we focused our experiments on ultrasound modality. To test the efficiency of our proposed method across other modalities, we performed experiments on two different public datasets.
\\

\textbf{GLAS Dataset} GLAnd Segmentation (GLAS) datatset contains microscopic images of Hematoxylin and Eosin (H\&E) stained slides and the corresponding ground truth annotations by expert pathologists. It contains a total of 165 images which are split into 85 images for training and 80 for testing. Since the images in the dataset are of different sizes, we resize every image to a resolution of $128\times128$ for all our experiments. We compare the performance of our proposed KiU-Net with leading state-of-the-art methods Seg-Net \cite{badrinarayanan2017segnet} and U-Net \cite{ronneberger2015u}. Table \ref{glasquan} shows the quantitative results for GLAS dataset, where KiU-Net achieves a 4\% improvement in terms of dice accuracy over U-Net. It can be noted that there were no pre-processing or post-processing steps that were used for any of these experiments. Fig \ref{Fig:exp} illustrates the qualitative results of the methods. It is visible from the images that our method captures edges better and gives a better segmentation prediction than the compared methods.

\begin{table}
	\centering
	\label{glasquan}
	\caption{Quantitative results for GLAS Dataset.}
	\begin{tabular}{l|c|c}
		\hline
		Method       & DICE Accuracy (in \%)  & Jaccard Index   \\ \hline 
		Seg-Net \cite{badrinarayanan2017segnet}       &  78.61  & 65.96     \\
		U-Net \cite{ronneberger2015u}         & 79.76   & 67.63    \\ 
		
		KiU-Net (ours) & \textbf{83.25 }  & \textbf{72.78}   \\ \hline
		
	\end{tabular}

\end{table}

\begin{figure}[htbp]
	\begin{center}
		\centering
		\includegraphics[width=0.15\textwidth,height = 0.15\textwidth]{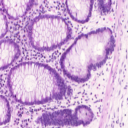}
		\includegraphics[width=0.15\textwidth,height = 0.15\textwidth]{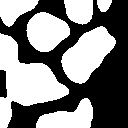}
		\includegraphics[width=0.15\textwidth,height = 0.15\textwidth]{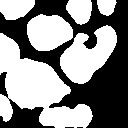}
		\includegraphics[width=0.15\textwidth,height = 0.15\textwidth]{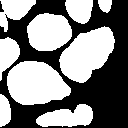}
		\includegraphics[width=0.15\textwidth,height = 0.15\textwidth]{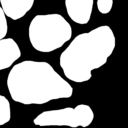}
		
		\vskip4pt
		\includegraphics[width=0.15\textwidth,height = 0.15\textwidth]{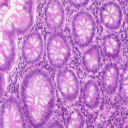}
		\includegraphics[width=0.15\textwidth,height = 0.15\textwidth]{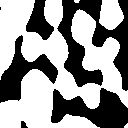}
		\includegraphics[width=0.15\textwidth,height = 0.15\textwidth]{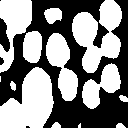}
		\includegraphics[width=0.15\textwidth,height = 0.15\textwidth]{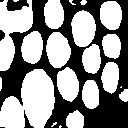}
		\includegraphics[width=0.15\textwidth,height = 0.15\textwidth]{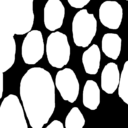}
		
		\vskip4pt
		
		(a)\hskip45pt(b)\hskip45pt(c)\hskip45pt(d)\hskip45pt(e)\\
		\vskip -10pt
		
		\caption{Qualitative results on sample test images. (a) H\& E stained input image. Predictions by (b) Seg-Net \cite{badrinarayanan2017segnet}, (c) U-Net \cite{ronneberger2015u} (d) KiU-Net (ours) and (e) Ground Truth}
		\label{Fig:exp}
	\end{center}
\end{figure}

\begin{figure}[htbp]
	\begin{center}
		\centering
		\includegraphics[width=0.15\textwidth,height = 0.15\textwidth]{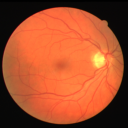}
		\includegraphics[width=0.15\textwidth,height = 0.15\textwidth]{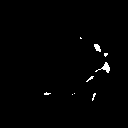}
		\includegraphics[width=0.15\textwidth,height = 0.15\textwidth]{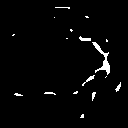}
		\includegraphics[width=0.15\textwidth,height = 0.15\textwidth]{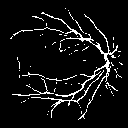}
		\includegraphics[width=0.15\textwidth,height = 0.15\textwidth]{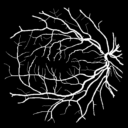}
		
		\vskip4pt
		\includegraphics[width=0.15\textwidth,height = 0.15\textwidth]{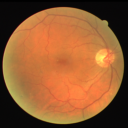}
		\includegraphics[width=0.15\textwidth,height = 0.15\textwidth]{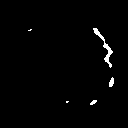}
		\includegraphics[width=0.15\textwidth,height = 0.15\textwidth]{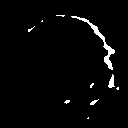}
		\includegraphics[width=0.15\textwidth,height = 0.15\textwidth]{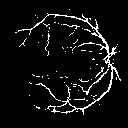}
		\includegraphics[width=0.15\textwidth,height = 0.15\textwidth]{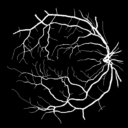}
		
		\vskip4pt
		
		(a)\hskip45pt(b)\hskip45pt(c)\hskip45pt(d)\hskip45pt(e)\\
		\vskip -10pt
		
		\caption{Qualitative results on sample test images. (a) H\& E stained input image. Predictions by (b) Seg-Net \cite{badrinarayanan2017segnet}, (c) U-Net \cite{ronneberger2015u} (d) KiU-Net (ours) and (e) Ground Truth}
		\label{Fig:exp2}
	\end{center}
\end{figure}

\textbf{RITE Dataset} RITE (Retinal Images vessel Tree Extraction) is a dataset that contains segmentation of arteries and veins on retinal fundus images. The dataset contains 40 image sets split into 20 for training and 20 for testing. The fundus images come with a vessel reference standard, and a Arteries/Veins (A/V) reference standard. For our experiments, we train our networks to predict the vessel segmentation from the fundus input images. We resize all the images to $128\times 128$ for our experiments. We compare the performance of our proposed KiU-Net with leading state-of-the-art methods Seg-Net \cite{badrinarayanan2017segnet} and U-Net \cite{ronneberger2015u}. Table \ref{glasquan} shows the quantitative results for RITE dataset, where KiU-Net achieves  significant improvement in terms of dice accuracy over U-Net. Fig \ref{Fig:exp2} illustrates the qualitative comparisons. It should be noted here that the quality of results can be increased by following some pre-processing steps specific to fundus images. We did not use any specific pre-processing steps or specific loss functions. We conduct this experiment to just show the superiority of our proposed method over other methods.

\begin{table}
	\centering
	\label{fund}
	\caption{Quantitative results for RITE Dataset.}
	\begin{tabular}{l|c|c}
		\hline
		Method       & DICE Accuracy (in \%)  & Jaccard Index   \\ \hline 
		Seg-Net \cite{badrinarayanan2017segnet}       &  52.23  & 39.14     \\
		U-Net \cite{ronneberger2015u}         & 55.24   & 31.11    \\ 
		
		KiU-Net (ours) & \textbf{75.17 }  & \textbf{60.37}   \\ \hline
		
	\end{tabular}

\end{table}

\section*{Network Architecture}

Tables \ref{tab1} and \ref{tab2} show the configuration of the Ki-Net and U-Net branch of our KiU Network. $H$ and $W$ are 128 each for the ultrasound train and test images. 
\begin{table}[htp]
	\begin{center}
		\centering
		\caption{Configuration of the Ki-Net branch of KiU-Net.}
		
		\resizebox{1\textwidth}{!}{
			\begin{tabular}{c|c|c|c|c|c|c}
				\hline \hline
				Block name                          & Layer     & Kernel size/Scale Factor& Filters & Padding & Input size & Output size \\ \hline
				\multirow{9}{*}{Encoder}           & Conv1     & 3 $\times$ 3       & 32      & 1          & 1 $\times$ H $\times$ W  & 32 $\times$ H $\times$ W  \\ \cline{2-7} 
				& Upsampling     & 2 $\times$ 2       & -      & -          & 32 $\times$ H $\times$ W & 32 $\times$ 2H $\times$ 2W  \\ \cline{2-7} 
				& ReLU     & -      & -       & -          & 32 $\times$ 2H $\times$ 2W & 32 $\times$ 2H $\times$ 2W   \\ \cline{2-7}
				& Conv2    & 3 $\times$ 3       & 64      & 1          & 32 $\times$ 2H $\times$ 2W  & 64 $\times$ 2H $\times$ 2W  \\ \cline{2-7} 
				& Upsampling     & 2 $\times$ 2       & -      & -          & 64 $\times$ 2H $\times$ 2W & 64 $\times$ 4H $\times$ 4W  \\ \cline{2-7} 
				& ReLU     & -       & -       & -          & 64 $\times$ 4H $\times$ 4W & 64 $\times$ 4H $\times$ 4W   \\ \cline{2-7}
				& Conv3     & 3 $\times$ 3        & 128      & 1          & 64 $\times$ 4H $\times$ 4W  & 128 $\times$ 4H $\times$ 4W  \\ \cline{2-7} 
				& Upsampling     & 2 $\times$ 2       & -      & -          & 128 $\times$ 4H $\times$ 4W & 2C $\times$ 8H $\times$ 8W  \\ \cline{2-7} 
				& ReLU     & -       & -       & -          & 128 $\times$ 8H $\times$ 8W & 128 $\times$ 8H $\times$ 8W   \\ \hline					
				\multirow{9}{*}{Decoder}           & Conv1     & 3 $\times$ 3       & 128      & 1          & 128 $\times$ 8H $\times$ 8W  & 128 $\times$ 8H $\times$ 8W  \\ \cline{2-7} 
				& Max-Pooling     & 2 $\times$ 2       & -      & -          & 128 $\times$ 8H $\times$ 8W & 128 $\times$ 4H $\times$ 4W  \\ \cline{2-7} 
				& ReLU     & -      & -       & -          & 128 $\times$ 4H $\times$ 4W & 128 $\times$ 4H $\times$ 4W   \\ \cline{2-7}
				& Conv2    & 3 $\times$ 3       & 64      & 1          & 128 $\times$ 4H $\times$ 4W  & 128 $\times$ 4H $\times$ 4W  \\ \cline{2-7} 
				& Max-Pooling      & 2 $\times$ 2       & -      & -          & 64 $\times$ 4H $\times$ 4W & 64 $\times$ 2H $\times$ 2W  \\ \cline{2-7} 
				& ReLU     & -       & -       & -          & 64 $\times$ 2H $\times$ 2W & 64 $\times$ 2H $\times$ 2W   \\ \cline{2-7}
				& Conv3     & 3 $\times$ 3        & 32      & 1          & 64 $\times$ 2H $\times$ 2W  & 32 $\times$ 2H $\times$ 2W  \\ \cline{2-7} 
				& Max-Pooling      & 2 $\times$ 2       & -      & -          & 32 $\times$ 2H $\times$ 2W & 32 $\times$ H $\times$ W  \\ \cline{2-7} 
				& ReLU     & -       & -       & -          & 32 $\times$ H $\times$ W & 32 $\times$ H $\times$ W   \\ \hline

			\end{tabular}
		}
		\label{tab1}
	\end{center}
\end{table}

\begin{table}[htp!]
	\begin{center}
		\centering
		\caption{Configuration of the U-Net branch of KiU-Net.}
		\label{tab2}
		\resizebox{1\textwidth}{!}{
			\begin{tabular}{c|c|c|c|c|c|c}
				\hline \hline
				Block name                          & Layer     & Kernel size/Scale Factor& Filters & Padding & Input size & Output size \\ \hline

				\multirow{9}{*}{Encoder}           & Conv1     & 3 $\times$ 3       & 32      & 1          & 1 $\times$ H $\times$ W  & 32 $\times$ H $\times$ W  \\ \cline{2-7} 
				& MaxPooling     & 2 $\times$ 2       & -      & -          & 32 $\times$ H $\times$ W & 32 $\times$ H/2 $\times$ W/2  \\ \cline{2-7} 
				& ReLU     & -      & -       & -          & 32 $\times$ H/2 $\times$ W/2 & 32 $\times$ H/2 $\times$ W/2   \\ \cline{2-7}
				& Conv2    & 3 $\times$ 3       & 64      & 1          & 32 $\times$ H/2 $\times$ W/2  & 64 $\times$ H/2 $\times$ W/2  \\ \cline{2-7} 
				& MaxPooling     & 2 $\times$ 2       & -      & -          & 64 $\times$ H/2 $\times$ W/2 & 64 $\times$ H/4 $\times$ W/4  \\ \cline{2-7} 
				& ReLU     & -       & -       & -          & 64 $\times$ H/4 $\times$ W/4 & 64 $\times$ H/4 $\times$ W/4   \\ \cline{2-7}
				& Conv3     & 3 $\times$ 3        & 128      & 1          & 64 $\times$ H/4 $\times$ W/4  & 128 $\times$ H/4 $\times$ W/4  \\ \cline{2-7} 
				& MaxPooling     & 2 $\times$ 2       & -      & -          & 128 $\times$ H/4 $\times$ W/4 & 128 $\times$ H/8 $\times$ W/8  \\ \cline{2-7} 
				& ReLU     & -       & -       & -          & 128 $\times$ H/8 $\times$ W/8 & 128 $\times$ H/8 $\times$ W/8      \\ \hline						
				
				\multirow{9}{*}{Decoder}           & Conv1     & 3 $\times$ 3       & 128      & 1          & 512 $\times$ H/32 $\times$ W/32& 128 $\times$ H/32 $\times$ W/32   \\ \cline{2-7} 
				& Upsampling     & 2 $\times$ 2       & -      & -          & 128 $\times$ H/32 $\times$ W/32  & 128 $\times$ H/16 $\times$ W/16  \\ \cline{2-7} 
				& ReLU     & -      & -       & -          & 128 $\times$ H/16 $\times$ W/16 & 128 $\times$ H/16 $\times$ W/16  \\ \cline{2-7}
				& Conv2    & 3 $\times$ 3       & 64      & 1          & 128 $\times$ H/16 $\times$ W/16 & 64 $\times$ H/16 $\times$ W/16  \\ \cline{2-7} 
				& Upsampling     & 2 $\times$ 2       & -      & -          & 64 $\times$ H/16 $\times$ W/16 & 64 $\times$ H/8 $\times$ W/8  \\ \cline{2-7} 
				& ReLU     & -       & -       & -          & 64 $\times$ H/8 $\times$ W/8 & 64 $\times$ H/8 $\times$ W/8   \\ \cline{2-7}
				& Conv3     & 3 $\times$ 3        & 32      & 1          & 64 $\times$ H/8 $\times$ W/8 & 32 $\times$ H/8 $\times$ W/8  \\ \cline{2-7} 
				& Upsampling     & 2 $\times$ 2       & -      & -          &  32 $\times$ H/8 $\times$ W/8 &  32 $\times$ H/4 $\times$ W/4  \\ \cline{2-7} 
				& ReLU     & -       & -       & -          & 32 $\times$ H/4 $\times$ W/4 & 32 $\times$ H/4 $\times$ W/4   				
				\\ \hline	
			\end{tabular}
		}
	\end{center}
\end{table}

\section*{Acknowledgement}
This work was supported by the NSF grant 1910141.

\bibliographystyle{splncs04}
\bibliography{oc_MICCAI20}

\end{document}